\documentclass[twocolumn]{aastex631}
\usepackage{amsmath}
\usepackage{stmaryrd}
\usepackage{amssymb}
\usepackage{wasysym}

\newcommand{\RNum}[1]{\uppercase\expandafter{\romannumeral #1\relax}}

\begin{document}
\title{Multiband Optical Variability of the Blazar 3C 454.3 on Diverse Timescales}

\author[0009-0007-3214-602X]{Karan Dogra}\email{karandogra987@gmail.com}
\affiliation{Aryabhatta Research Institute of Observational Sciences(ARIES), Manora Peak, Nainital 263001, India}
\affiliation{Department of Applied Physics, Mahatma Jyotiba Phule Rohilkhand University, Bareilly 243006, India}

\author[0000-0002-9331-4388]{Alok C. Gupta}\email{acgupta30@gmail.com}
\affiliation{Aryabhatta Research Institute of Observational Sciences(ARIES), Manora Peak, Nainital 263001, India}
\affiliation{Xinjiang Astronomical Observatory, Chinese Academy of Sciences, 150 Science-1 Street, Urumqi 830011, China}

\author[0000-0003-1784-2784]{C. M. Raiteri}
\affiliation{INAF, Osservatorio Astrofisico di Torino, via Osservatorio 20, I-10025 Pino Torinese, Italy}

\author[0000-0003-1743-6946]{M. Villata}
\affiliation{INAF, Osservatorio Astrofisico di Torino, via Osservatorio 20, I-10025 Pino Torinese, Italy}

\author[0000-0002-1029-3746]{Paul J. Wiita}
\affiliation{Department of Physics, The College of New Jersey, PO Box 7718, Ewing, NJ 08628, USA}

\author[0000-0002-0319-5873]{S. O. Kurtanidze}
\affiliation{Abastumani Observatory, Mt. Kanobili, 0301 Abastumani, Georgia}

\author[0000-0001-6158-1708]{S. G. Jorstad}
\affiliation{Institute for Astrophysical Research, Boston University, 725 Commonwealth Avenue, Boston, MA 02215, USA}
\affiliation{Astronomical Institute, Saint Petersburg State University, 7/9 Universitetskaya nab., St. Petersburg, 199034, Russia}

\author[0000-0002-0766-864X]{R. Bachev}
\affiliation{Institute of Astronomy and National Astronomical Observatory, Bulgarian Academy of Sciences, 72 Tsarigradsko shosse Blvd., 1784 Sofia, Bulgaria}

\author[0000-0002-6710-6868]{G. Damljanovic}
\affiliation{Astronomical Observatory, Volgina 7, 11060 Belgrade, Serbia}

\author{C. Lorey}
\affiliation{Hans-Haffner-Sternwarte, Naturwissenschaftliches Labor f\"{u}r Sch\"{u}ler am FKG, Friedrich-Koenig-Gymnasium, D-97082 W$\ddot{u}$rzburg, Germany}

\author[0000-0003-4147-3851]{S. S. Savchenko}
\affiliation{Astronomical Institute, Saint Petersburg State University, 7/9 Universitetskaya nab., St. Petersburg, 199034, Russia}
\affiliation{Special Astrophysical Observatory, Russian Academy of Sciences, 369167, Nizhnii Arkhyz, Russia}
\affiliation{Pulkovo Observatory, St.Petersburg, 196140, Russia}

\author[0009-0008-5761-3701]{O. Vince}
\affiliation{Astronomical Observatory, Volgina 7, 11060 Belgrade, Serbia}

\author{M. Abdelkareem}
\affiliation{National Research Institute of Astronomy and Geophysics (NRIAG), 11421 Helwan, Cairo, Egypt}

\author{F. J. Aceituno}
\affiliation{Instituto de Astrof\'{i}sica de Andaluc\'{i}a, IAA-CSIC, Glorieta de la Astronom\'{i}a s/n, E-18008 Granada, Spain}

\author[0000-0002-0433-9656]{J. A. Acosta-Pulido}
\affiliation{Instituto de Astrof\'{i}sica de Canarias (IAC), E-38200 La Laguna, Tenerife, Spain}
\affiliation{Universidad de La Laguna, Departamento de Astrof\'{i}sica, E-38206 La Laguna, Tenerife, Spain}

\author[0000-0002-3777-6182]{I. Agudo}
\affiliation{Instituto de Astrof\'{i}sica de Andaluc\'{i}a, IAA-CSIC, Glorieta de la Astronom\'{i}a s/n, E-18008 Granada, Spain}

\author[0000-0001-5125-6397]{G. Andreuzzi}
\affiliation{INAF, TNG Fundaci{\'o}n Galileo Galilei, La Palma, E-38712, Spain}

\author{S. A. Ata}
\affiliation{National Research Institute of Astronomy and Geophysics (NRIAG), 11421 Helwan, Cairo, Egypt}

\author{G. V. Baida}
\affiliation{Crimean Astrophysical Observatory of the Russian Academy of Sciences, P/O Nauchny 298409, Crimea}

\author{L. Barbieri}
\affiliation{Orciatico Astronomical Observatory, Orciatico (Pisa), Italy}

\author[0000-0003-0611-5784]{D. A. Blinov}
\affiliation{Institute of Astrophysics, Foundation for Research and Technology - Hellas, Voutes, 70013 Heraklion, Greece}
\affiliation{Department of Physics, University of Crete, 71003, Heraklion, Greece}

\author{G. Bonnoli}
\affiliation{Instituto de Astrof\'{i}sica de Andaluc\'{i}a, IAA-CSIC, Glorieta de la Astronom\'{i}a s/n, E-18008 Granada, Spain}
\affiliation{INAF Osservatorio Astronomico di Brera, Via E. Bianchi 46, 23807 Merate (LC), Italy}

\author[0000-0002-7262-6710]{G. A. Borman}
\affiliation{Crimean Astrophysical Observatory of the Russian Academy of Sciences,
P/O Nauchny 298409, Russia}

\author[0000-0001-5843-5515]{M. I. Carnerero}
\affiliation{INAF, Osservatorio Astrofisico di Torino, via Osservatorio 20, I-10025 Pino Torinese, Italy}

\author[0000-0001-5252-1068]{D. Carosati}
\affiliation{EPT Observatories, Tijarafe, La Palma, E-38780, Spain}
\affiliation{INAF, TNG Fundaci{\'o}n Galileo Galilei, La Palma, E-38712, Spain}

\author{V. Casanova}
\affiliation{Instituto de Astrof\'{i}sica de Andaluc\'{i}a, IAA-CSIC, Glorieta de la Astronom\'{i}a s/n, E-18008 Granada, Spain}

\author[0000-0003-0262-272X]{W. P. Chen}
\affiliation{Institute of Astronomy, National Central University, Taoyuan 32001, Taiwan}

\author[0000-0003-0721-5509]{Lang Cui}
\affiliation{Xinjiang Astronomical Observatory, Chinese Academy of Sciences, 150 Science-1 Street, Urumqi 830011, China}

\author[0000-0002-9751-8089]{E. G. Elhosseiny}
\affiliation{National Research Institute of Astronomy and Geophysics (NRIAG), 11421 Helwan, Cairo, Egypt}

\author{D. Elsaesser}
\affiliation{Hans-Haffner-Sternwarte, Naturwissenschaftliches Labor f\"{u}r Sch\"{u}ler am FKG, Friedrich-Koenig-Gymnasium, D-97082 W$\ddot{u}$rzburg, Germany}
\affiliation{Astroteilchenphysik, TU Dortmund, Otto-Hahn-Str. 4A, D-44227 Dortmund, Germany}

\author{J. Escudero}
\affiliation{Instituto de Astrof\'{i}sica de Andaluc\'{i}a, IAA-CSIC, Glorieta de la Astronom\'{i}a s/n, E-18008 Granada, Spain}

\author{M. Feige}
\affiliation{Hans-Haffner-Sternwarte, Naturwissenschaftliches Labor f\"{u}r Sch\"{u}ler am FKG, Friedrich-Koenig-Gymnasium, D-97082 W$\ddot{u}$rzburg, Germany}

\author[0000-0002-8855-3923]{K. Gazeas}
\affiliation{Section of Astrophysics, Astronomy and Mechanics, Department of Physics, National and Kapodistrian University of Athens, GR-15784 Zografos, Athens, Greece}

\author{L. E. Gennadievna}
\affiliation{Astronomical Institute, Saint Petersburg State University, 7/9 Universitetskaya nab., St. Petersburg, 199034, Russia}

\author{T. S. Grishina}
\affiliation{Astronomical Institute, Saint Petersburg State University, 7/9 Universitetskaya nab., St. Petersburg, 199034, Russia}

\author[0000-0002-4455-6946]{Minfeng Gu}
\affiliation{Shanghai Astronomical Observatory, Chinese Academy of Sciences, 80 Nandan Road, Shanghai 200030, China}

\author{V. A. Hagen-Thorn}
\affiliation{Astronomical Institute, Saint Petersburg State University, 7/9 Universitetskaya nab., St. Petersburg, 199034, Russia}

\author{F. Hemrich}
\affiliation{Hans-Haffner-Sternwarte, Naturwissenschaftliches Labor f\"{u}r Sch\"{u}ler am FKG, Friedrich-Koenig-Gymnasium, D-97082 W$\ddot{u}$rzburg, Germany}

\author{H. Y. Hsiao}
\affiliation{Institute of Astronomy, National Central University, Taoyuan 32001, Taiwan}

\author{M. Ismail}
\affiliation{National Research Institute of Astronomy and Geophysics (NRIAG), 11421 Helwan, Cairo, Egypt}

\author[0009-0005-7297-8985]{R. Z. Ivanidze}
\affiliation{Abastumani Observatory, Mt. Kanobili, 0301 Abastumani, Georgia}

\author[0000-0003-4298-3247]{M. D. Jovanovic}
\affiliation{Astronomical Observatory, Volgina 7, 11060 Belgrade, Serbia}

\author{T. M. Kamel}
\affiliation{National Research Institute of Astronomy and Geophysics (NRIAG), 11421 Helwan, Cairo, Egypt}

\author[0000-0002-5684-2114]{G. N. Kimeridze}
\affiliation{Abastumani Observatory, Mt. Kanobili, 0301 Abastumani, Georgia}

\author{E. N. Kopatskaya}
\affiliation{Astronomical Institute, Saint Petersburg State University, 7/9 Universitetskaya nab., St. Petersburg, 199034, Russia}

\author{D. Kuberek}
\affiliation{Hans-Haffner-Sternwarte, Naturwissenschaftliches Labor f\"{u}r Sch\"{u}ler am FKG, Friedrich-Koenig-Gymnasium, D-97082 W$\ddot{u}$rzburg, Germany}

\author[0000-0001-5385-0576]{O. M. Kurtanidze}
\affiliation{Abastumani Observatory, Mt. Kanobili, 0301 Abastumani, Georgia}
\affiliation{Engelhardt Astronomical Observatory, Kazan Federal University, Tatarstan, Russia}

\author{A. Kurtenkov}
\affiliation{Institute of Astronomy and National Astronomical Observatory, Bulgarian Academy of Sciences, 72 Tsarigradsko shosse Blvd., 1784 Sofia, Bulgaria}

\author{V. M. Larionov}
\affiliation{Astronomical Institute, Saint Petersburg State University, 7/9 Universitetskaya nab., St. Petersburg, 199034, Russia}

\author{L. V. Larionova}
\affiliation{Astronomical Institute, Saint Petersburg State University, 7/9 Universitetskaya nab., St. Petersburg, 199034, Russia}

\author[0000-0002-9137-7019]{M. Liao}
\affiliation{National Astronomical Observatories, Chinese Academy of Sciences, 20A Datun Road, Chaoyang District, Beijing 100101, China}
\affiliation{Chinese Academy of Sciences South America Center of Astronomy, National Astronomical Observatories, CAS, Beijing 100101, China}
\affiliation{Instituto de Estudios Astrofisicos Facultad de Ingenieria y Ciencias Universidad Diego Portales Av. Ej$\acute{e}$rcito 441, Santiago, Chile}

\author{H. C. Lin}
\affiliation{Institute of Astronomy, National Central University, Taoyuan 32001, Taiwan}

\author{K. Mannheim}
\affiliation{Hans-Haffner-Sternwarte, Naturwissenschaftliches Labor f\"{u}r Sch\"{u}ler am FKG, Friedrich-Koenig-Gymnasium, D-97082 W$\ddot{u}$rzburg, Germany}
\affiliation{Lehrstuhl f$\ddot{u}$r Astronomie, Universit$\ddot{a}$t W$\ddot{u}$rzburg,  D-97074 W$\ddot{u}$rzburg, Germany}

\author[0000-0003-3779-6762]{A. Marchini}
\affiliation{University of Siena, Astronomical Observatory, Via Roma 56. 53100, Siena, Italy}

\author[0000-0002-3596-4307]{C. Marinelli}
\affiliation{Dip. di Scienze Fisiche, della Terra e dell’Ambiente, Universit\'{a} di Siena, Via Roma 56, 53100, Siena, Italy}

\author[0000-0001-7396-3332]{A. P. Marscher}
\affiliation{Institute for Astrophysical Research, Boston University, 725 Commonwealth Avenue, Boston, MA 02215, USA}

\author{D. Morcuende}
\affiliation{Instituto de Astrof\'{i}sica de Andaluc\'{i}a, IAA-CSIC, Glorieta de la Astronom\'{i}a s/n, E-18008 Granada, Spain}

\author{D. A. Morozova}
\affiliation{Astronomical Institute, Saint Petersburg State University, 7/9 Universitetskaya nab., St. Petersburg, 199034, Russia}

\author{S. V. Nazarov}
\affiliation{Crimean Astrophysical Observatory of the Russian Academy of Sciences, P/O Nauchny 298409, Russia}

\author[0000-0003-0408-7177]{M. G. Nikolashvili}
\affiliation{Abastumani Observatory, Mt. Kanobili, 0301 Abastumani, Georgia}

\author{D. Reinhart}
\affiliation{Hans-Haffner-Sternwarte, Naturwissenschaftliches Labor f\"{u}r Sch\"{u}ler am FKG, Friedrich-Koenig-Gymnasium, D-97082 W$\ddot{u}$rzburg, Germany}

\author{J. O. Santos}
\affiliation{Instituto de Astrof\'{i}sica de Andaluc\'{i}a, IAA-CSIC, Glorieta de la Astronom\'{i}a s/n, E-18008 Granada, Spain}

\author{A. Scherbantin}
\affiliation{Hans-Haffner-Sternwarte, Naturwissenschaftliches Labor f\"{u}r Sch\"{u}ler am FKG, Friedrich-Koenig-Gymnasium, D-97082 W$\ddot{u}$rzburg, Germany}
\affiliation{Lehrstuhl f$\ddot{u}$r Astronomie, Universit$\ddot{a}$t W$\ddot{u}$rzburg,  D-97074 W$\ddot{u}$rzburg, Germany}

\author[0000-0002-1839-3936]{E. Semkov}
\affiliation{Institute of Astronomy and National Astronomical Observatory, Bulgarian Academy of Sciences, 72 Tsarigradsko shosse Blvd., 1784 Sofia, Bulgaria}

\author{E. V. Shishkina}
\affiliation{Astronomical Institute, Saint Petersburg State University, 7/9 Universitetskaya nab., St. Petersburg, 199034, Russia}

\author[0000-0002-6985-2143]{L. A. Sigua}
\affiliation{Abastumani Observatory, Mt. Kanobili, 0301 Abastumani, Georgia}

\author{A. K. Singh}
\affiliation{Department of Applied Physics, Mahatma Jyotiba Phule Rohilkhand University, Bareilly 243006, India}

\author{A. Sota}
\affiliation{Instituto de Astrof\'{i}sica de Andaluc\'{i}a, IAA-CSIC, Glorieta de la Astronom\'{i}a s/n, E-18008 Granada, Spain}

\author{R. Steineke}
\affiliation{Hans-Haffner-Sternwarte, Naturwissenschaftliches Labor f\"{u}r Sch\"{u}ler am FKG, Friedrich-Koenig-Gymnasium, D-97082 W$\ddot{u}$rzburg, Germany}

\author[0000-0002-4105-7113]{M. Stojanovic}
\affiliation{Astronomical Observatory, Volgina 7, 11060 Belgrade, Serbia}

\author{A. Strigachev}
\affiliation{Institute of Astronomy and National Astronomical Observatory, Bulgarian Academy of Sciences, 72 Tsarigradsko shosse Blvd., 1784 Sofia, Bulgaria}

\author[0000-0003-1423-5516]{A. Takey}
\affiliation{National Research Institute of Astronomy and Geophysics (NRIAG), 11421 Helwan, Cairo, Egypt}

\author[0000-0002-8279-9236]{Amira A. Tawfeek}
\affiliation{National Research Institute of Astronomy and Geophysics (NRIAG), 11421 Helwan, Cairo, Egypt}

\author{I. S. Troitskiy}
\affiliation{Astronomical Institute, Saint Petersburg State University, 7/9 Universitetskaya nab., St. Petersburg, 199034, Russia}

\author{Y. V. Troitskaya}
\affiliation{Astronomical Institute, Saint Petersburg State University, 7/9 Universitetskaya nab., St. Petersburg, 199034, Russia}

\author[0000-0002-3211-4219]{An-Li Tsai}
\affiliation{Institute of Astronomy, National Central University, Taoyuan 32001, Taiwan}
\affiliation{Department of Physics, National Sun Yat-sen University, Kaohsiung, 80424, Taiwan}

\author{A. A. Vasilyev}
\affiliation{Astronomical Institute, Saint Petersburg State University, 7/9 Universitetskaya nab., St. Petersburg, 199034, Russia}

\author[0009-0002-7669-7425]{K. Vrontaki}
\affiliation{Section of Astrophysics, Astronomy and Mechanics, Department of Physics, National and Kapodistrian University of Athens, GR-15784 Zografos, Athens, Greece}

\author[0000-0002-8366-3373]{Zhongli Zhang}
\affiliation{Shanghai Astronomical Observatory, Chinese Academy of Sciences, 80 Nandan Road, Shanghai 200030, China}
\affiliation{Key Laboratory of Radio Astronomy and Technology, Chinese Academy of Sciences, A20 Datun Road, Chaoyang District, Beijing 100101, China}

\author{A. V. Zhovtan}
\affiliation{Crimean Astrophysical Observatory of the Russian Academy of Sciences, P/O Nauchny 298409, Crimea}

\author{N. Zottmann}
\affiliation{Hans-Haffner-Sternwarte, Naturwissenschaftliches Labor f\"{u}r Sch\"{u}ler am FKG, Friedrich-Koenig-Gymnasium, D-97082 W$\ddot{u}$rzburg, Germany}

\author[0000-0002-4521-6281]{Wenwen Zuo}
\affiliation{Shanghai Astronomical Observatory, Chinese Academy of Sciences, 80 Nandan Road, Shanghai 200030, China}




\begin{abstract}
Due to its peculiar and highly variable nature, the blazar 3C 454.3 has been extensively monitored by the WEBT team. Here, we present for the first time these long-term optical flux and color variability results using data acquired in B, V, R, and I bands over a time span of $\sim$ 2 decades. We include data from WEBT collaborators and public archives such as SMARTS, Steward Observatory, and ZTF. The data are binned and segmented to study the source over this long term when more regular sampling was available. During our study, the long-term spectral variability reveals a redder when brighter (RWB) trend, which, however, stabilizes at a particular brightness cutoff $\sim$ 14.5 mag in the I-band, after which it saturates and evolves into a complex state. This trend indicates increasing jet emission dominance over accretion disk emission until jet emission completely dominates. Plots of the spectral index variation (following $F_{\nu} \propto \nu^{-\alpha}$) reveal a bimodal distribution using a one-day binning. These correlate with two extreme phases of 3C 454.3, an outburst or high flux state and quiescent or low flux state, which are respectively jet and accretion disk dominated. We have also conducted intra-day variability studies of nine light curves and found that six of them are variable. Discrete Correlation Function (DCF) analysis between different optical waveband pairs peak at zero lags, indicating co-spatial emission in different optical bands.
\end{abstract}


\section{Introduction}

Blazars, a distinct subclass of radio-loud (RL) Active Galactic Nuclei (AGN), exhibit flux changes over the whole electromagnetic (EM) spectrum, covering time intervals from minutes to several years. This category of AGN is further divided into BL Lacertae objects (BL Lacs) and Flat-Spectrum Radio Quasars (FSRQs). BL Lacs have essentially non-existent or very weak emission lines (equivalent widths $\leq$ 5\AA ) in their UV (ultraviolet) to optical spectra \citep{1991ApJS...76..813S,1996MNRAS.281..425M}, whereas FSRQs are characterized by significant emission lines. \citep{1978PhyS...17..265B,1997A&A...327...61G}. The multi-wavelength (MW) spectral energy distribution (SED) of blazars is double-hump shaped \citep[e.g.][]{1995ApJ...440..525V,1998MNRAS.299..433F}. According to the location of the first hump peak, blazars have been more recently classified into three sub-classes: low synchrotron frequency peaked (LSP) blazars, intermediate synchrotron frequency peaked (ISP) blazars, and high synchrotron frequency peaked (HSP) blazars. For LSPs, the synchrotron peak frequency ($\nu_{sy}$)  of the first hump peak is $\nu_{sy} \leq 10^{14}$ Hz; for ISPs, it is $10^{14} \rm{Hz} \leq \nu_{sy} \leq 10^{15}$ Hz; and HSPs have $\nu_{sy} \geq 10^{15}$ Hz \citep{2010ApJ...716...30A}. The first hump of the SED is due to the synchrotron emission from the electrons in the relativistic jet. The second hump is widely accepted as coming from the inverse Compton (IC) scattering of either the photons emitted by the synchrotron process responsible for the first hump (synchrotron self-Compton; SSC) or of photons emerging from the accretion disk (AD), emission lines or dusty torus (external Compton; EC) \citep[e.g.][]{1997ARA&A..35..445U,2007Ap&SS.309...95B}. Aside from these leptonic models, the second hump may be explained by hadronic models, which involve emission from relativistic protons or muon synchrotron radiation \citep{2003APh....18..593M}. 
\\
\\
Blazar jets are oriented nearly along the viewer's line of sight, resulting in the detected radiation being significantly Doppler-enhanced \citep[e.g.][]{1995PASP..107..803U}. The majority of blazar emissions across the EM spectrum are non-thermal. But in FSRQs, a bump is sometimes seen to rise above them from $\sim$1 $\mu$m and cutoff below 0.6 KeV, which is called the big-blue bump (BBB) \citep{1978Natur.272..706S,1990A&ARv...2..125B}. Two emission mechanisms have been seriously proposed for the BBB: optically thin thermal free-free radiation \citep[e.g.][]{1990ApJ...363L..17A,1990ApJ...363L..21F,1993ApJ...412..513B,1996A&A...314..393C} and optically thick thermal radiation from the accretion disk  
\citep[e.g.][]{1978Natur.272..706S,1982ApJ...254...22M,1987ApJ...321..305C,1989MNRAS.238..897L}.\\
\\
It has been shown over the past few decades that blazars exhibit variations in flux, spectrum, and polarization in all EM bands that are accessible on time spanning a couple of minutes to many years \citep[e.g.][]{2015MNRAS.451.3882A,2016MNRAS.458.1127G,2017MNRAS.472..788G,2024MNRAS.527.1344D}. Blazar variability on timescales ranging from a few minutes to less than a day is referred to as micro-variability \citep{1989Natur.337..627M}, or intraday variability (IDV) \citep{1995ARA&A..33..163W}. Blazar variability on time scales ranging from a few days to several weeks is referred to as short-term variability (STV), while variability over months to decades is referred to as long-term variability (LTV) \citep{2004A&A...422..505G}. Most of the variability shown by blazars in the EM spectrum is characterized by non-linearity, stochasticity, and a lack of periodicity \citep[e.g.][]{2017ApJ...849..138K}. \\
\\
Earlier studies reveal that, in general, there are two types of spectral variations trends  in blazars, with BL Lacs often showing bluer when brighter (BWB) behavior \citep[e.g.][]{2006MNRAS.366.1337S,2011JApA...32...87G,2012MNRAS.425.3002G}  
and FSRQs frequently becoming redder When brighter (RWB) \citep[e.g][]{2006A&A...450...39G,2012ApJ...756...13B} trends. Nonetheless, some studies have found opposite trends, i.e., FSRQs following BWB  \citep{2011JApA...32...87G} and BL Lacs following RWB. Some previous studies also suggested that the long-term color variability of BL Lac objects have no definite trend \citep{2002A&A...390..407V,2004A&A...421..103V,2006MNRAS.366.1337S,2021MNRAS.501.1100R,2023MNRAS.522..102R}, or achromatic behavior. A study conducted by \cite{2015RAA....15.1784Z}, using a sample of 49 FSRQs and 22 BL Lac objects from the SMARTS monitoring program, found that 35 FSRQs and 11 BL Lacs respectively followed RWB and BWB trends. The majority of these studies lack a sufficient number of quasi-simultaneous data points or have been conducted over a short period of time. This motivated us to study the dichotomous color behavior of blazars along with their flux variability over diverse timescales. We wanted to investigate if there is a long-term trend in color with flux that can be well studied with an extensive data set or if a smooth transition is possible in color-magnitude space. \\

The FSRQ 3C 454.3, situated at a red-shift of 0.859 \citep{1991MNRAS.250..414J}, is one of the most rigorously monitored blazars over the whole EM spectrum on different timescales for the study of its various properties using flux, spectral, and polarization data \citep[e.g.][and references therein]{2006A&A...445L...1F,2006A&A...453..817V,2009A&A...504L...9V,2007A&A...473..819R,2008A&A...485L..17R,2008A&A...491..755R,2011A&A...534A..87R,2007MNRAS.382L..82G,2009ApJ...697L..81B,2009ApJ...699..817A,2010PASJ...62..645S,2014ApJ...784..141S,2010ApJ...715..362J,2013ApJ...773..147J,2011A&A...528L..10B,2011ApJ...736L..38V,2011ApJS..195...19O,2012AJ....143...23G,2012ApJ...758...72W,2013ApJ...763L..36L,2013ApJ...779..100I,2017Galax...5....3V,2017MNRAS.464.2046K,2017MNRAS.472..788G,2019A&A...631A...4N,2019ApJ...887..185S,2021ApJ...906....5A,2021ApJS..253...10F,2021MNRAS.504.5074S}. Following the launch of the {\tt Fermi} satellite, it was identified as one of the most luminous $\gamma$-ray emitting blazar \citep{2010ApJ...721.1383A}. For this object, the supermassive black hole (SMBH) mass has been estimated to be in the range of (0.5 -- 2.3) $\times \rm{10}^{9} \ \rm{M}_{\odot}$  \citep[e.g.][and references therein]{2017MNRAS.472..788G,2019A&A...631A...4N}. This study focuses on the flux and color variability of FSRQ 3C 454.3, IDV analysis of the source on multiple nights in the R-band, along with some of the long-term studies discussed in Section 3. The color or spectral variability studies provide insight into the currently dominant emission source, i.e. the jet or the accretion disk, of the object under study. For the first time, the extensive observations of this source discussed here include multiple outbursts and quiescent phases, providing a rare opportunity to study the flux and color variations of a blazar over a long baseline of $\sim$ 20 years. \\

In Section 2, we describe the extensive optical data we employ from many observatories around the world, along with archival data. In Section 3, we present our analysis methods and findings related to IDV, STV, and LTV studies. Section 4 discusses our results from this extensive study with Section 5 summarizing our conclusions.

\begin{table*}
\caption{ List of optical data sets from different observatories with aperture size and filters used.}
\begin{tabular}{llllccccc}
\hline
\hline
Observatory & MJD Start & MJD Stop  & Origin  & Aperture & Filters & Data points &  Marker & Color \\
&&&&(cm)&&$(N_{obs})$&&\\
\hline
\hline
Abastumani&58501.7&59875.8&Georgia                        &70&R& 488  &        $\ast$        &Tan\\
ASV$^{\emph{a}}$&58723.8&58818.8&Serbia                   &140/60&BVRI& 83 &     $\square$             &Black\\      
Athens$^{\emph{b}}$&57736.6&58793.8&Greece                &40&R& 55  &     $\circ$             &Cyan\\
Belogradchik&59603.7&59872.0&Bulgaria                     &60&VRI& 12   &        $\triangledown$         &Lime\\
Crimean (AP-7)&55401.0&59544.7&Russia                     &70&BVRI& 1656   &         $\triangleleft$         &Blue\\
Crimean (ST-7)&53516.0&59848.9&Russia                     &70&BVRI& 1445  &           $\triangleright$       &Light sea-green\\
Hans Haffner&57280.8&60097.0&Germany                      &50&BVR& 794   &          $\square$      &Red\\
KAO$^{\emph{c}}$&59576.7&59969.7&Egypt                    &188&BVR& 174 &      $\Delta$             &Maroon\\
Lulin (SLT)&58251.8&60073.8&Taiwan                        &40&VR&  824   &         $\triangleright$       &Indigo\\
MAPCAT$^{\emph{d}}$&59456.5&59938.3&Spain                 &220&R&  14 &        $\diamond$         &Gold\\       
Perkins&55295.5&59930.1&United States                     &180&BVRI& 1763  &         $\Delta$        &Green\\
Rozhen&55864.7&59107.7&Bulgaria                           &200&BVRI& 185   &        $\pentagon$         &Lime\\
Rozhen$^{\emph{e}}$&59581.7&59870.9&Bulgaria              &50/70&BVRI& 16  &      $\pentagon$             &Lime\\
Siena&58380.9&58400.8&Italy                               &30&R&  11    &      $\triangledown$         &Slategrey\\
Skinakas&56079.0&58709.9&Greece                           &130&BVRI& 666  &        $\square$          &Lime\\
SNO$^{\emph{f}}$&59476.4&59958.3&Spain                    &90&R& 127  &      $\hexagon$            &Gold\\
St. Petersburg&53965.8&59901.7&Russia                     &40&BVRI& 1101  &           $\varhexagon$         &Olive\\   
Teide$^{\emph{g}}$&55669.5&60065.5&Spain                  &Various&R& 330 & + &Cyan\\
Tijarafe&57617.0&59952.8&Spain                            &40&R& 924 & x                  &Violet\\
\hline
SMARTS$^{\emph{h}}$&54662.3&57964.3&Chile                   &130/150&BVR& 2178  &     $\circ$             &Tomato\\
Steward$^{\emph{i}}$&54745.2&58306.4&United States                   &154/230&VR& 1101  &         +         &Orange\\
WEBT$^{\emph{j}}$&53965.8&55543.8&International                  &Various&BVRI& 15017  &     $\diamond$  &Purple\\
ZTF$^{\emph{k}}$&58254.4&59966.1&United States                   &122&gr&  881   &        $\varhexagon$        &Maroon\\      
\hline 
\label{table:1}
\end{tabular} \\
\centering
Notes.\\
$^a$\emph{Astronomical station Vidojevica}\\
$^b$\emph{University of Athens Observatory (UOAO)}\\
$^c$\emph{Kottamia Astronomical Observatory}\\
$^d$\emph{Calar ALTO Observatory}\\
$^e$\emph{Rozhen 50/70 cm Schmidt telescope}\\
$^f$\emph{Sierra Nevada Observatory}\\
$^g$\emph{Teide observatory with STELLA-I, IAC80 and Las Cumbres facilities}\\
$^h$\emph{Small and Medium Aperture Research Telescope System with 130 cm and 150 cm telescopes}\\
$^i$\emph{Steward Observatory Support of the Fermi Mission with 154 cm Kuiper and 230 cm Bok telescope}\\
$^j$\emph{Whole Earth Blazar Telescope Archive}\\ 
$^k$\emph{Zwicky Transient Facility}\\
\end{table*}

\section{Observation AND DATA PROCESSING} 
In the current work, monitoring of the FSRQ 3C 454.3 from many different telescopes across the globe made between June 2004 to June 2023 are combined. Some observation details are mentioned in  \autoref{table:1}. All the telescopes in \autoref{table:1} utilize CCD detectors and broadband Johnson-Cousin filters (UBVRI); however, only the BVRI data are used in our analysis. Most of the data has been provided by the Whole Earth Blazar Telescope (WEBT\footnote{\url{https://www.oato.inaf.it/blazars/webt/}}), which is an international team of observers who perform persistent optical-to-radio observations of a sample of $\gamma$-loud blazars, especially during enhanced activity phases. WEBT data for this paper come both from the WEBT archive, containing results from past observing campaigns on the source \citep{2006A&A...453..817V,2007A&A...464L...5V,2007A&A...473..819R,2008A&A...485L..17R,2008A&A...491..755R,2009A&A...504L...9V,2011A&A...534A..87R,2011ApJ...736L..38V}, and from more recent observations by the teams listed in \autoref{table:1}.\\

\begin{figure*}
    \centering
    \includegraphics[width=17cm,height=17cm]{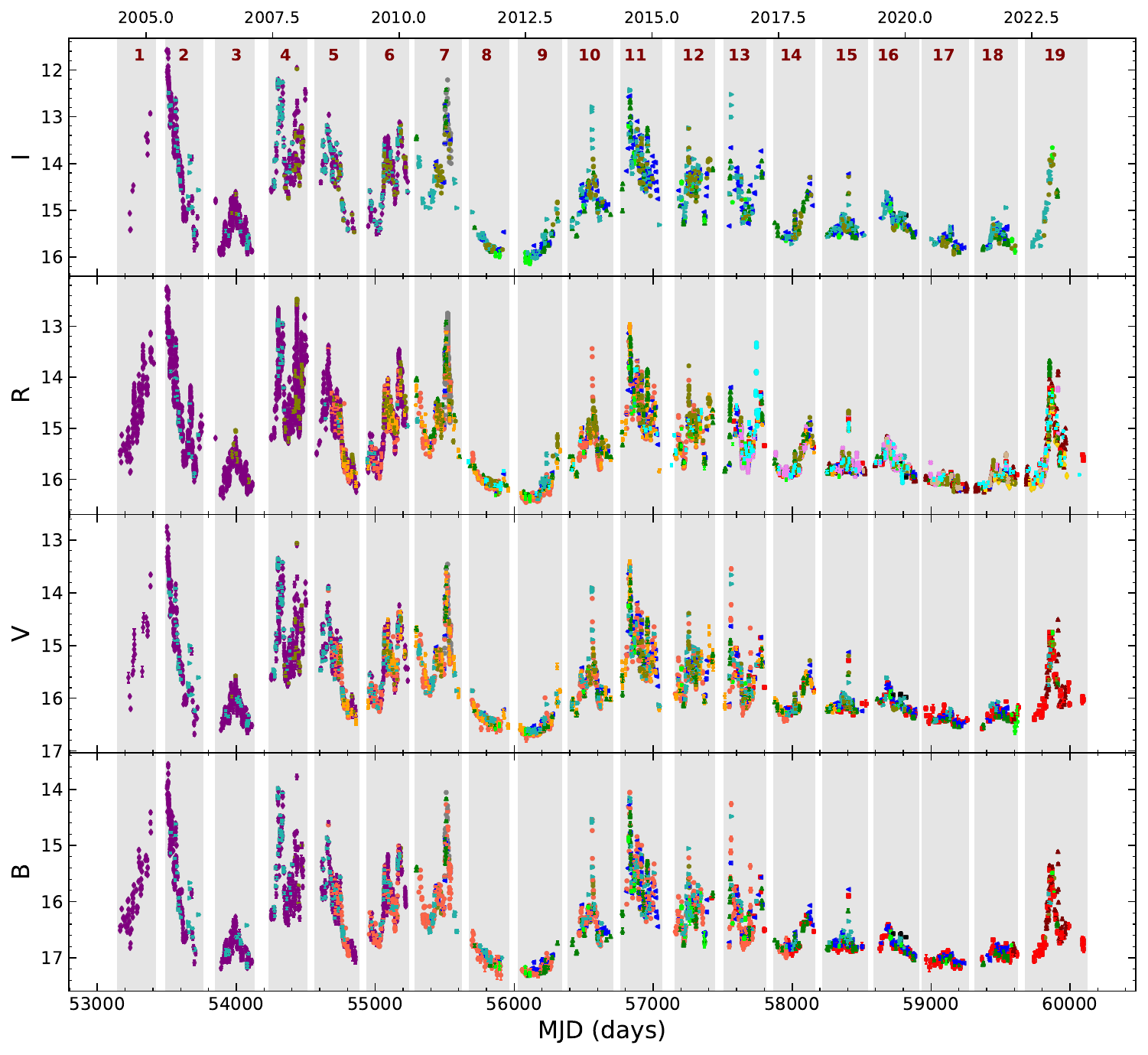}
    \caption{Extensively monitored optical LCs of the source in BVRI bands. The light curve consists of data from public archives and WEBT collaborators, identified by the colors given in Table 1.}
    \label{fig:1}
\end{figure*}

\noindent Data from public archives have also been incorporated.  These are the Steward Observatory Support of the Fermi Mission \citep{2009arXiv0912.3621S} (Steward\footnote{\url{https://james.as.arizona.edu/~psmith/Fermi/}}), Small and Medium Aperture Research Telescope System  (SMARTS\footnote{\url{http://www.astro.yale.edu/smarts/glast/home.php}}) \citep{2012ApJ...756...13B}, and Zwicky Transient Facility (ZTF\footnote{\url{https://www.ztf.caltech.edu/ztf-public-releases.html}}) Observatory \citep{2019PASP..131a8002B}. 
Steward Observatory provided data in the V and R filters, SMARTS in B, V, and R filters, and ZTF in g and r filters. \\

As shown in \autoref{fig:1}, the entire light curve is divided into 19 segments (Seg) reflecting the observation cycle gaps. As the light curve involves data from many telescopes around the globe, offset corrections are made if required. The WEBT archive data is taken as the reference for this as it is already offset corrected. The optical photometry data from Steward, SMARTS, and ZTF observatories, along with unpublished data from current WEBT collaborators, is fully reduced and calibrated.
The data are generally cleaned and reduced using the standard techniques discussed below. The first step is the cleaning of the raw image files using readily available techniques in Image Reduction and Analysis Facility {\tt (IRAF)\footnote{IRAF is distributed by the National Optical Astronomy Observatories, which are operated by the Association of Universities for Research in Astronomy, Inc., under a cooperative agreement with the National Science Foundation}}  package. For this reduction, we make a master bias frame by median combining all the bias frames using the {\tt zero-combine} method for the individual night, which then is subtracted from science and flat-fielding images. After that, the bias-subtracted flat frames from each night are median combined using the {\tt flatcombine} method to produce master flat frames for individual filters. These master flat frames are then used to normalize the science frames to remove pixel-to-pixel variations due to varying sensitivity and the presence of dust on filters and/or CCDs, or both. Next, cosmic ray removal was carried out for the bias subtracted and flat fielded science frames using the {\tt cosmicrays} package. After the cleaning process, the instrumental magnitudes of FSRQ 3C 454.3 and the standard stars are usually extracted using {\tt DAOPHOT\footnote{Dominion Astronomical Observatory Photometry}} \RNum{2} \citep{1987PASP...99..191S,1992JRASC..86...71S} software using aperture photometry. \\
\\
We performed aperture photometry on all the science frames using four concentric apertures with radii of one, two, three, and four times the FWHM. Several studies \citep{2016MNRAS.458.1127G,2019ApJ...871..192P} indicate that the signal-to-noise ratio is usually best for the two times FWHM  aperture, so we have generally employed it. The data reduction for the Bulgarian and Georgian telescopes is discussed in \cite{2012AJ....143...23G} whereas \cite{2020MNRAS.496.1430P} discusses the same for the Serbian telescopes.\\

\begin{table*}
\centering
\caption{Short-term segment-wise variability amplitudes of 3C 454.3 }
\begin{tabular}{cccccccccccccccccccccccc}
\hline
\hline
\vspace{0.1cm}
& Band & $V_F$  &   & Band & $V_F$  &   & Band & $V_F$& & Band & $V_F$  &   & Band & $V_F$  &   & Band & $V_F$ \\
&  &  &  &  &  &  &  &  &  &  &  &  &  &   &  &  &  \\
\hline
\hline
      &B&1.23&                   &B&2.40&                  &B&2.24&                &B&3.89&                    &B&1.64&                   &B&0.51\\
Seg &V&1.36&         Seg     &V&2.29&          Seg   &V&2.61&        Seg  &V&3.75&         Seg      &V&1.69&          Seg   &V&0.61\\
 1     &R&1.32&       4            &R&2.51&     7             &R&2.20&10                &R&3.99&13                    &R&1.69&16                   &R&0.68\\
      &I&1.49&                   &I&2.31&                  &I&2.82&                &I&4.17&                    &I&1.75&                   &I&0.87\\
\hline
       &B&3.33&                   &B&2.18&                   &B&0.41&               &B&3.38&                    &B&0.86&                   &B&0.20\\
Seg  &V&3.08&        Seg      &V&1.88&          Seg    &V&0.31&      Seg   &V&2.89&         Seg      &V&1.04&          Seg   &V&0.25\\
 2      &R&3.41&      5             &R&2.02&     8              &R&0.51&11               &R&2.85&14                    &R&1.29&17                   &R&0.34\\
       &I&3.53&                   &I&1.99&                   &I&0.52&               &I&2.88&                    &I&1.68&                   &I&0.53\\
\hline
      &B&0.81&                   &B&1.52&                  &B&0.31&                &B&1.93&                   &B&1.42&                    &B&0.37\\
Seg &V&0.91&         Seg     &V&1.53&          Seg   &V&0.50&       Seg    &V&2.01&         Seg    &V&1.53&           Seg    &V&0.40\\
 3     &R&1.08&       6            &R&1.69&     9             &R&0.48&12                &R&1.89&15                   &R&1.72&18                    &R&0.40\\
      &I&1.25&                   &I&1.87&                  &I&0.58&                &I&2.55&                   &I&2.02&                    &I&0.56\\
\hline
\label{tab:2}
\end{tabular}
\end{table*} 

For magnitude calibration and variability analysis, the standards\footnote{\url{https://www.oato.inaf.it/blazars/webt/3c-454-3-2251158/}} proposed by the GLAST-AGILE Support Program (GASP), i.e., stars 1, 2, 3, and 4 are observed in all science frames. Among these, star 4 has been used to calibrate the instrumental magnitudes of the source as its magnitude and color are closer to it, which reduces errors while performing differential photometry. Since all the standard stars and  3C 454.3 have been observed under the same weather conditions and air mass, there is no need to perform atmospheric extinction corrections.

\section{Analysis techniques and results} 
The variability of an AGN over the broad EM spectrum is one of its defining characteristics. This variability is an excellent tool for probing the physical properties of the emission regions. Inspection of \autoref{fig:1} shows that 3C 454.3 is clearly variable on long-term and short-term timescales, and we compute the amplitude of variability during each segment. To investigate the presence of IDV for the blazar 3C 454.3, we have employed the enhanced F-test and Nested ANOVA test techniques, which offer greater reliability than other widely used techniques such as the F-test. Discrete correlation function analysis has been implemented to quantify any possible correlation among optical bands.

\subsection{Amplitude variation}
For the light curves identified as being variable, we determined the amplitude of variation using the formula provided by \citep{1996A&A...305...42H}:
\begin{equation}
    V_F = \frac{\sqrt{(F_{max}-F_{min})^2 - 2\sigma^2}}{\bar{F}}
\end{equation}
Where $F_{max}$, $F_{min}$, and $\bar{F}$ are the highest, lowest, and mean fluxes of the variable light curve, with $\sigma$ being the average error. The resulting STV amplitude variations in respective optical bands are shown in \autoref{tab:2} for the first 18 segments. Segments 2, 10, and 11 are the most variable, while segments 16, 17, and 18 are the least variable across the entire light curve.

\subsection{Color variation and model selection}
\begin{figure*}
    \centering
    \includegraphics[width=0.90\textwidth]{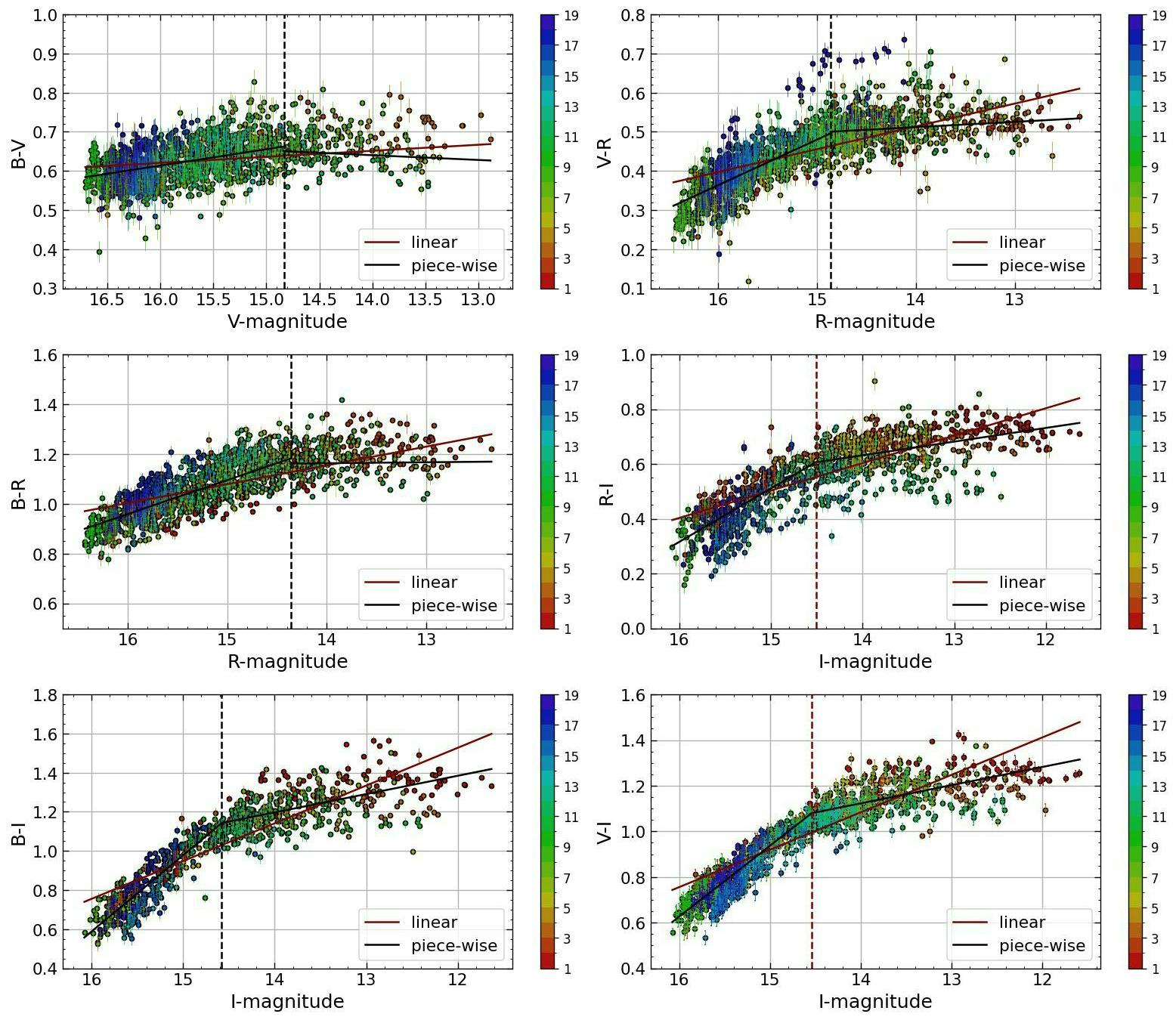}
    \caption{Color index vs magnitude plots}
    \label{fig:2}
\end{figure*}

To study the spectral variation, we plotted color index vs magnitude diagrams in \autoref{fig:2} in which data points are color-coded according to which segment of the light curve contributed points to the color-magnitude (CM) diagram. The plots are arranged with increasing frequency differences between the bands (from top to bottom). All the plots indicate a redder when brighter trend up to a cutoff magnitude (shown by a maroon vertical dashed line), although the trend is clearer and much more localized in the color-magnitude space when the bands are well separated, which can be understood in terms of the emission from the jet and the disk. The thermal part is mainly ascribed to the accretion disk, whereas the non-thermal component arises from relativistic jet emissions. Nonetheless, a strict linear trend is not seen, as the color indices saturate at lower magnitudes (brighter phases), i.e., they increase more slowly and eventually flatten in the long term as the source brightens. This flattening was first noted by  \cite{2006A&A...453..817V}.\\
\\
To verify and quantify this, two functions have been fitted to all the plots in \autoref{fig:2}, namely linear and piece-wise, with their forms given below:
\begin{equation}
    y = m x + c ~,
\end{equation}
and 
\begin{equation}
\begin{split}
    y_{1} &= m_{1} x + c_{1}~, \\
    y_{2} &= m_{2} x + c_{2}~,\\
\end{split}
\end{equation}
where $m$ is the slope of the line and $c$ is the intercept. The parameters have been estimated using the Maximum Likelihood Estimation (MLE), which is implemented using {\tt emcee\footnote{\url{https://emcee.readthedocs.io/en/stable/}}} package. As the name suggests, MLE tries to maximize the likelihood function to get the best estimates of the parameters; however, maximizing the likelihood is similar to minimizing the negative logarithm of the likelihood, which brings an additional benefit, as the logarithm of the likelihood transforms the computations from multiplicative to summation series which are much easier to work with. If the error bars are correct, Gaussian, and independent, the results of MLE would be the same as those of the simple least squares methodology. Still, since those conditions do not necessarily apply, we can optimize the probability distribution function (PDF) using MLE if we consider that some fractional amount underestimates the error bars. This correction makes our log-likelihood function become:
\begin{equation}
\begin{split}
    ln(L) &= -\frac{1}{2}\sum_{n}\left[{\frac{(y_{n}-mx_{n}-c)^2}{s_{n}^2} + ln(2\pi s_{n}^2) }\right] \\
    {\rm and}\; s_{n}^2 &= \sigma^{2} + f^2(mx+c)^2, ~\\
\end{split}
\end{equation}
\\

\begin{table*}
\centering
\caption{IDV analysis results for nested ANOVA and enhanced F-test}
\begin{tabular}{cccccccccccc}
\hline\hline
Obs. date & Band & \multicolumn{4}{c}{Enhanced F-test} & \multicolumn{4}{c}{Nested ANOVA test} & Status & $V_F$  \\
yyyy-mm-dd& & D.O.F($\nu_1$,$\nu_2$) & $F_{enh}$ & $F_{c}$&$p-value$ & D.O.F($\nu_1$,$\nu_2$) & F & $F_{c}$& $p-value$ & &$(\%)$ \\
\hline\hline
2022-10-25&R&374,1122&3.34&1.21&1.11x$10^{-16}$&14,360&262.82&2.13&1.11x$10^{-16}$&V&7.33\\
2022-10-31&R&131,524&10.94&1.36&1.11x$10^{-16}$&10,121&208.60&2.47&1.11x$10^{-16}$&V&6.60\\
2022-11-14&R&434,2170&2.52&1.18&1.11x$10^{-16}$&28,406&24.71&1.77&1.11x$10^{-16}$&V&6.75\\
2022-12-21&R&164,492&2.49&1.33&1.08x$10^{-14}$&14,150&5.43&2.20&2.42x$10^{-08}$&V&16.41\\
2022-12-23&R&89,267&3.25&1.47&8.44x$10^{-14}$&8,81&7.38&2.74&2.50x$10^{-07}$&V&25.34\\
2022-12-25&R&155,465&3.50&1.34&1.11x$10^{-16}$&12,143&11.18&2.31&1.55x$10^{-15}$&V&14.11\\
2023-10-07&R&66,198&0.54&1.56&0.997&10,55&2.68&2.66&0.009&NV&-\\
2023-10-14&R&122,366&0.64&1.39&0.998&10,110&0.51&2.49&0.879&NV&-\\
2023-12-21&R&37,111&0.22&1.8&0.999&8,27&0.47&3.26&0.86&NV&-\\
\hline
\label{tab:5}
\end{tabular}
\end{table*}

where \emph{f} is the fractional amount by which variance is underestimated. If the errors assume a Gaussian distribution, \emph{f} is set to zero. A Markov chain Monte Carlo (MCMC) technique has been implemented to calculate the uncertainties in the evaluated parameters, which the MLE technique does not provide.
\\
\\
Now, to select between the models discussed above, Akaike's Information Criteria (AIC) and Bayesian Information Criteria (BIC) have been calculated using the log-likelihood calculated previously. AIC and BIC are derived using frequentist and Bayesian approaches, respectively, and are given by:
\begin{equation}
\begin{split}
    AIC &=2k - 2ln(L)~ ,\\
    BIC &=kln(n) - 2ln(L)~ ,\\
\end{split}
\end{equation}
where $k$, $n$, and $L$ represent the free parameters, data points, and likelihood function of the model, respectively. Both information criteria rely not only on how well the model fits the data but also consider the complexity of the respective fitting function. If we keep increasing the number of parameters in the model, the data will be over-fitted, a problem both information criteria penalize, although this penalty is higher in BIC than AIC. The model with lower values of AIC and BIC is preferred. Piece-wise linear functions fit the data better, as can be seen in \autoref{tab:3}; slope 1 and slope 2 represent the part of the fit to the left and right of the break-point, respectively. In \autoref{fig:2}, the maroon-dashed lines indicate the break-points of the piece-wise linear functions.

\subsection{Power enhanced F-test}
To check the intraday variability of FSRQ 3C 454.3, we have implemented the power-enhanced F-test as described by \citet{2014AJ....148...93D} and \citet{2015AJ....150...44D}. This test has been widely used in recent variability studies \citep[e.g.][]{2015MNRAS.452.4263G,2016MNRAS.460.3950P,2017MNRAS.466.2679K,2019ApJ...871..192P,2020MNRAS.496.1430P}. In this test, the instrumental blazar differential light curve (DLC) variance is compared with the combined instrumental DLC variances of comparison stars. The differential light curves are produced with respect to a reference star. The comparison stars which do not show variability among themselves are considered. The reference star is the one whose magnitude is closest to the source's. The test statistics are given by: 
\begin{equation}
    F=\frac{s_{blz}^2}{s_c^2} ~,
\end{equation}
where $s_{blz}^2$ is the variance of the target DLC, and $s_c^2$ is the combined variance of the DLCs of the comparison stars.\\
\\

The value of $s_c^2$ is given by:
\begin{equation}
        s_c^2 =\frac{\sum_{j=1}^{k}\sum_{i=1}^{N_i}s_{j,i}^2}{(\sum_{j=1}^{k}N_j) - k} ~.
\end{equation}
Here, $N_{j}$ is the number of observations of the $j$th comparison star, and $k$ is the number of comparison stars. The scaled square deviation of the $j$th comparison star is
\begin{equation}
    s_{j,i}^2 = \omega_j(m_{j,i} - \bar m_j)^2~,
\end{equation}
\noindent where $\omega_{j}$ is the scaling factor, $m_{j,i}$ is the $i$th differential magnitude of the $j$th star and $\bar m_{j}$ is the mean magnitude of the $j$th star. The scaling factor is defined as the ratio of the average square error of the blazar DLC to the average square error of the $j$th comparison star DLC.\\
\begin{table*}
\centering
\caption{Fitting parameters for the color-magnitude plots}
\begin{tabular}{ccccc}
\hline\hline
     Color versus magnitude & Fitting function & Slope and Break-point & Akaike's information & Bayesian information\\
     &&&criteria&criteria\\
\hline\hline
     &Linear&slope = $-0.0154 \pm 0.0004  $&18501.78&18512.71  \\
     B-V vs V&Piece-wise linear&slope 1 = $ -0.0417 \pm 0.0007  $&16386.47&16408.33  \\
     &&slope 2 = $ 0.0121 \pm 0.0011 $&&  \\
     &&break-point = $ 14.833 \pm 0.007$&&  \\
\hline
     &Linear&slope = $ -0.0583\pm0.0003  $&16864.10&16875.24  \\
     V-R vs R&Piece-wise linear&slope 1 =$-0.1164 \pm 0.0010 $&8525.66&8547.93  \\
     &&slope 2 = $  -0.0136\pm0.0007  $&&  \\
     &&break-point = $ 14.862 \pm0.010  $&&  \\
\hline
     &Linear&slope = $-0.0756 \pm 0.0003$&52233.52&52244.38  \\
     B-R vs R&Piece-wise linear&slope 1 = $-0.1363 \pm 0.0007 $&38710.39& 38732.10 \\
     &&slope 2 = $ -0.0037\pm0.0011 $&&  \\
     &&break-point = $ 14.362\pm 0.016$&&  \\
\hline
     &Linear&slope = $ -0.1002\pm0.0004 $&22229.64&22239.80  \\
     R-I vs I&Piece-wise linear&slope 1 = $ -0.1954\pm0.0017 $&13553.37&13573.71  \\
     &&slope 2 = $ -0.0500\pm 0.0007$&&  \\
     &&break-point = $ 14.509\pm0.015 $&&  \\
\hline
     &Linear&slope = $-0.1933 \pm0.0005 $&49477.65&49487.39  \\
     B-I vs I&Piece-wise linear&slope 1 = $-0.3904 \pm0.0019 $ &23843.73&23863.22  \\
     &&slope 2 = $-0.0937 \pm0.0009 $&&  \\
     &&break-point = $14.577 \pm 0.012$&&  \\
\hline
     &Linear&slope = $-0.1639 \pm0.0004 $&34557.59&34567.89  \\
     V-I vs I&Piece-wise linear&slope 1 = $-0.3107 \pm0.0015 $&13133.09&13153.70  \\
     &&slope 2 = $ -0.0797\pm0.0008 $&&  \\
     &&break-point = $14.538 \pm0.007 $&&  \\
\hline       
\label{tab:3}
\end{tabular}
\end{table*}

During the observations, we observed six field stars, out of which star 4 was chosen as the reference star, having the magnitude closest to the source. As we have the same number of observations ($M$) for field stars ($q$) and the source, the degrees of freedom for the numerator and denominator are $\nu_1 = M-1$ and $\nu_2 = q(M-1)$, respectively.
Along with the test statistic, which is the F-value, we need to calculate the critical F-value ($F_c$) along with the corresponding p-values. A comparison of $F$ with $F_c$ tells us about the significance of our result at that probability level, whereas the p-value indicates the probability of obtaining a result at least as extreme as the one observed, assuming the null hypothesis is correct.
Using these parameters, we have calculated the enhanced F-test ($F_{enh}$) value with the critical F-value ($F_{c}$) at a confidence level of 99\% or $\alpha=0.01$. Only when $F_{enh} \geq F_{c}$ do we label the source as variable. The results are given in \autoref{tab:5}. The light curves found to be variable are marked as V, and the non-variable ones are marked as NV. The amplitude of variability has also been computed for these LCs, with their results shown in the amplitude column of \autoref{tab:5}.

\begin{figure*}
    \centering
    \includegraphics[scale=0.62]{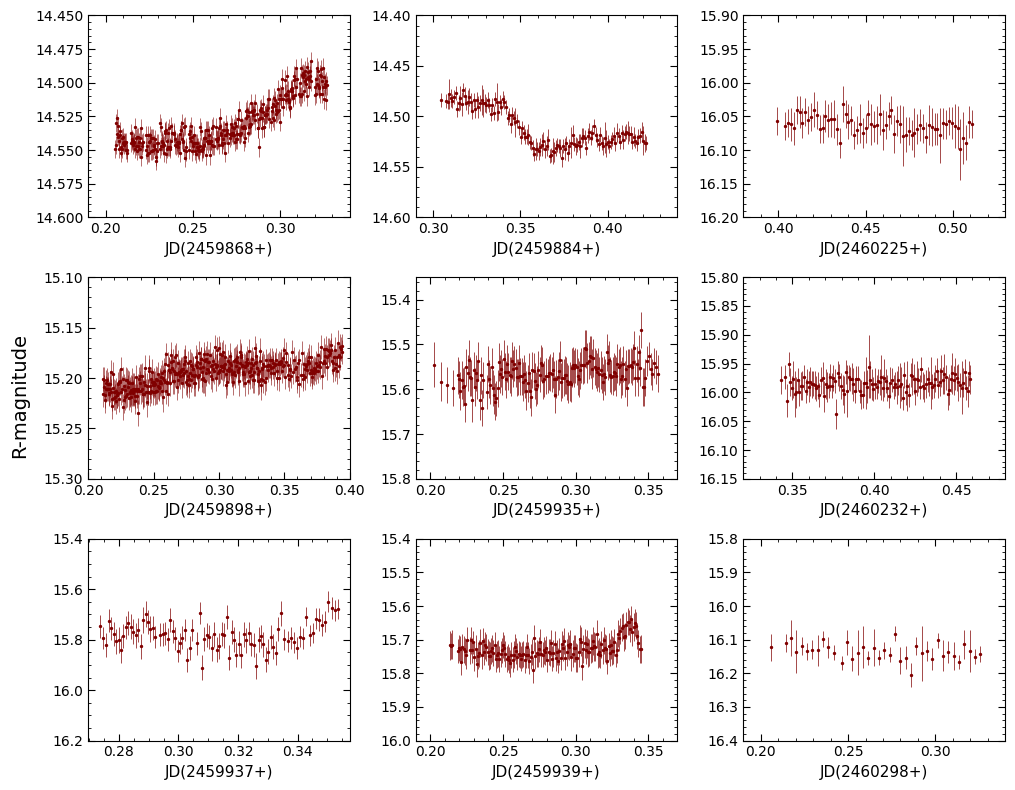}
    \caption{Intra-day LCs of FSRQ 3C 454.3 in R-filter for nine days. Each plot shows the respective date (in JD on the x-axis) on which observation is taken.}
    \label{fig:5}
\end{figure*}

\subsection{Nested ANOVA test}
An analysis of variance (ANOVA) test compares the means of different populations and checks whether the populations have intrinsic variation or are different. The nested ANOVA is a successor to the ANOVA test that further divides the groups into subgroups and then compares the means of these nested observations \citep{1998ApJ...501...69D,2015AJ....150...44D}. This test uses multiple field stars as a reference to produce the DLCs of the source. During the observations, we observed eight field stars, out of which we took four to six reference stars (depending on the night) to generate differential light curves of the blazar.\\
\\ 
The test statistic, which is used to verify there is a significant difference between the group means, is given by:
\begin{equation}
    \label{eq:9}
    F = \frac{MS_G}{MS_{WG}} ~,
\end{equation}
where $MS_G$ represents the sum of squares (SS) average among groups and $MS_{WG}$ represents the SS average due to nested observations. The formulas for $MS_G$ and $MS_{WG}$, respectively, are:
\begin{equation}
\begin{split}
    MS_{G} &= \sum_{i=1}^{n}\frac{(\bar{m_i}-\bar m)^2}{r-1}~;\\
MS_{WG} &= \sum_{i=1}^{n}\sum_{j=1}^{n}\frac{(\bar{m_{ij}} - \bar{m_i})^2}{r(t-1)}~.
\end{split}
\end{equation}

\begin{figure*}
    \centering
    \includegraphics[width=0.8\textwidth]{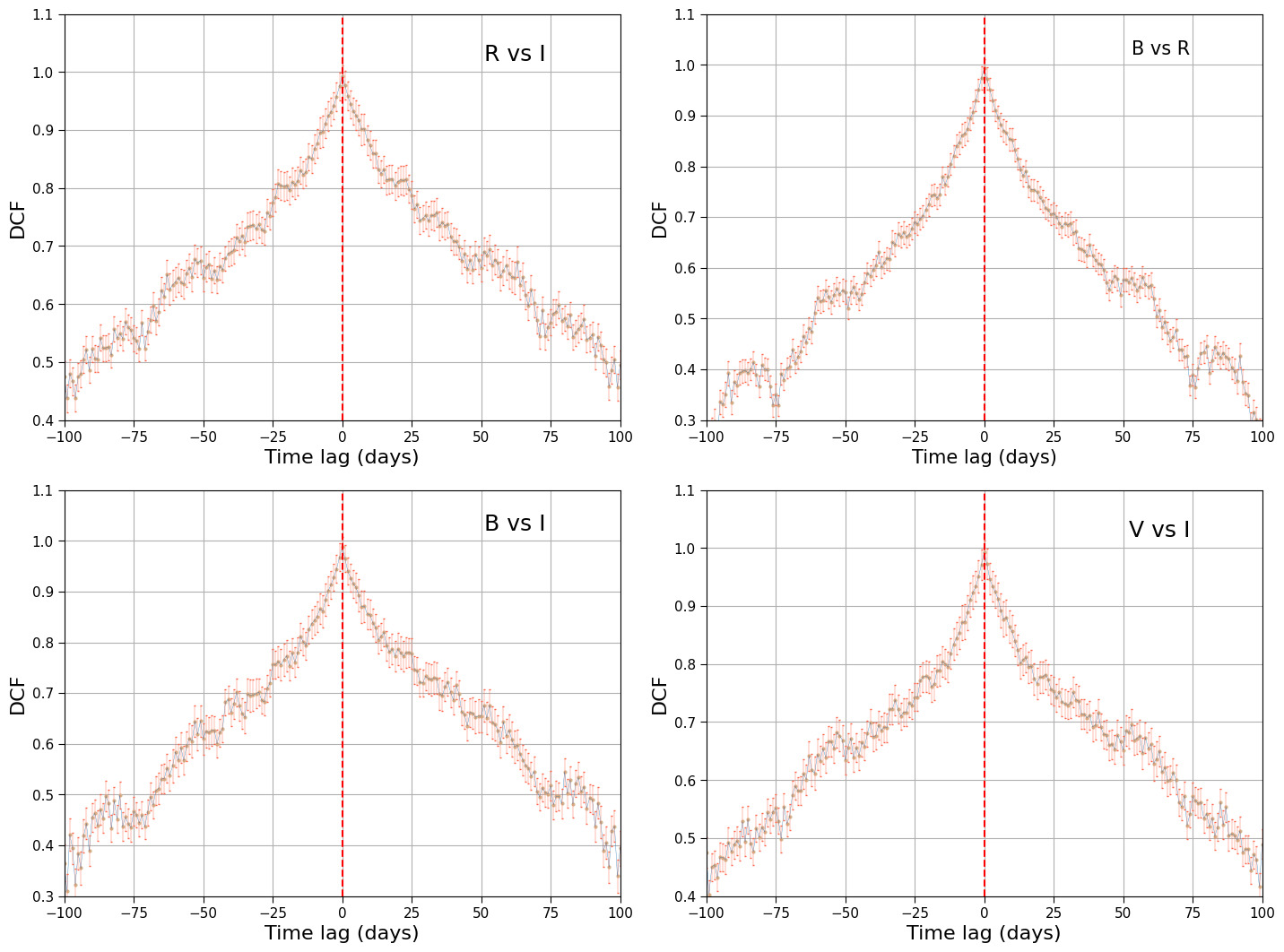}
    \caption{DCF plots for the entire observation period (here shown for 100 days lag) between different pairs of wavebands.}
    \label{fig:4}
\end{figure*}

Here, $a$ and $b$ represent the number of groups and the number of subgroups within a group, respectively.
The F-ratio in \autoref{eq:9} follows F-distribution with the numerator and denominator having $(r-1)$ and $r(t-1)$ degrees of freedom, respectively. If $F \geq F_c$, for $\alpha = 0.01$, then the source is taken to be variable, otherwise non-variable. 
Only if both the nested ANOVA and enhanced F-test detect statistically significant variations do we consider the source to be variable on the IDV time span. The nested ANOVA test results are given in \autoref{tab:5} with all the parameters already defined in the previous section. The first six light curves observed in the R-band are found to be variable, whereas the last three light curves are non-variable, as confirmed by both the test statistics being lower than the critical value.

\subsection{Discrete correlation function analysis}
The possible presence of different emission and absorbing regions around an AGN means that it is sometimes possible to find a time lag between different energy bands. Any time lag between different energy bands is usually investigated by calculating the discrete correlation function DCF proposed by \cite{1988ApJ...333..646E} and is generally used for data with irregular sampling. For a better estimation of errors associated with each bin, the method was further improved by \cite{1992ApJ...386..473H}. A positive DCF value indicates a possible positive correlation, whereas a DCF value of less than zero indicates an anti-correlation between the bands. \autoref{fig:4} shows DCF plots between different optical bands. The maximum in each DCF plot is shown by a red vertical dashed line, which lies at zero lag for all the plots. As there is no significant lag between the different band combinations, the optical emission can be taken to arise from a single region.

\begin{table}
\caption{Bimodal fit values to the spectral index distribution with $\mu, \sigma$ and $A$ representing mean, standard deviation and amplitude, respectively.}
\centering
\begin{tabular}{ccccccc}
    \hline\hline
    &$\mu_{1}$ &$\sigma_{1}$  &$A_{1}$  & $\mu_{2}$ & $\sigma_{2}$ & $A_{2}$\\
    \hline\hline
    value &-1.362  &0.192  & 67.430 &  -0.808&  0.169& 37.541\\
    error&0.011  &  0.012&  2.413&  0.016&  0.019&2.567\\ 
    \hline
\end{tabular}
\label{tab:bimodal}
\end{table}

\subsection{Spectral variation}
To study any variability of the optical spectrum, we used the fluxes in different filters and plotted the fluxes in these bands over the whole time series, yielding around 972 quasi-simultaneous (within one day) data points. \autoref{fig:5} shows some spectra at different times with straight lines showing the least square (LS) fits the data points. As the specific flux vs frequency plots shown in \autoref{fig:5} are in logarithmic space and follow $F_{\nu} \propto \nu^{-\alpha}$, where the slopes of the LS fits give optical spectral indices.
To correct for galactic extinction, we subtracted the extinction correction values (taken from NASA's Extra-galactic Database (NED)) of the respective band from our observations. The extinction corrections for the B, V, R, and I-band are 0.382, 0.289, 0.228, and 0.159, respectively. We used Vega as our standard star to convert calibrated magnitudes to flux values, which are given in SI units.  A temporal evolution of the fluxes at different bands has been generated with all the data points with a subset shown in \autoref{fig:5}. \\

\autoref{fig:7} shows the distribution of spectral indices, here defined as the slopes of the spectra shown in \autoref{fig:5}. 
This reveals a bimodal distribution, with the number of bins taken to be 34, and we fitted it using two Gaussian distributions, with their parameters given in \autoref{tab:bimodal}. The two Gaussian fits are shown by red dashed lines, and their combined fit is illustrated using a solid red line. This bimodal nature indicates the presence of two common phases for the states of 3C 454.3, i.e., a bright/outburst phase and a faint/quiescent phase, with the former contributing to the first hump and the latter contributing to the second hump of the bimodal distribution.

\section{discussion}

In this study, the optical photometric data of FSRQ 3C 454.3 collected for over 20 years have been analyzed. The data have been collected from the WEBT archive and its collaborators and include both new and published observations along with data from other public archives like SMARTS, Steward, and ZTF. During this long observation period, the object underwent a drastic change in its magnitude in most observing seasons, as shown in \autoref{fig:1}. The exceptions are segments 7, 8, 16, 17, and 18, which have a maximum magnitude change of $\sim$ 0.5 mag compared to 11 other segments, evincing a minimum change of $\sim$ 1 mag and reaching a maximum change of $\sim$ 4 mags in R-band in segment 8 and segment 2 respectively. As can be seen in \autoref{tab:2}, the longer wavelength bands are more variable than the shorter ones. This can be explained by considering that synchrotron emission is more variable than disk emission, and given that disk emission peaks near UV, it will make smaller contributions towards longer wavelengths, and hence, the variability will be strongest in the longer wavelength regime.
The diverse flux variability this blazar shows may be explained by fluctuations in the jet, along with the presence of the disk instabilities. For instance, the shock-in-jet model, in which internal shocks move along the jet, can explain the variability over diverse timescales \citep{1985ApJ...298..114M, 1995ARA&A..33..163W,2001MNRAS.325.1559S}. Some of the faster and small-scale amplitude variations can result from various instabilities in the accretion disk \citep{1992AIPC..254..251W,1993ApJ...411..602C}. 
Geometrical effects \citep[e.g.][]{1992A&A...255...59C} can also explain intraday variability \citep{1995ARA&A..33..163W} and the long-term variability via the production of fluctuations in the value of the Doppler factor that can result from changes in our line of sight to the predominant emission region in an inhomogeneous, curved, or twisting jet \citep{1992A&A...259..109G,1999A&A...347...30V,2017Natur.552..374R} along with misalignment of AGN jets \citep{2016ApJ...827...66S}. Another  scenario that can be responsible for jet fluctuations, especially during flaring events,  involves the collision of a blob with a quasi-stationary jet component \citep{2021ApJ...906....5A}.

The BWB trend observed in BL Lac objects can be explained by the fact that electrons accelerated by the shock front in the jet dissipate energy more rapidly, causing the higher energy bands to be more variable \citep{2002PASA...19..138M}. Alternatively, the increase in luminosity could be explained by the introduction of new electrons with a harder energy distribution compared to the previously cooled ones. On the other hand, in the case of FSRQs, the RWB trend can be simply explained by adding the redder jet emission to the bluer disk emission \citep{2006A&A...450...39G,2017ApJ...844..107I}, provided jet emission has not swamped the disk emission. \\

\begin{figure*}
\centering
    \includegraphics[width=0.78\textwidth]{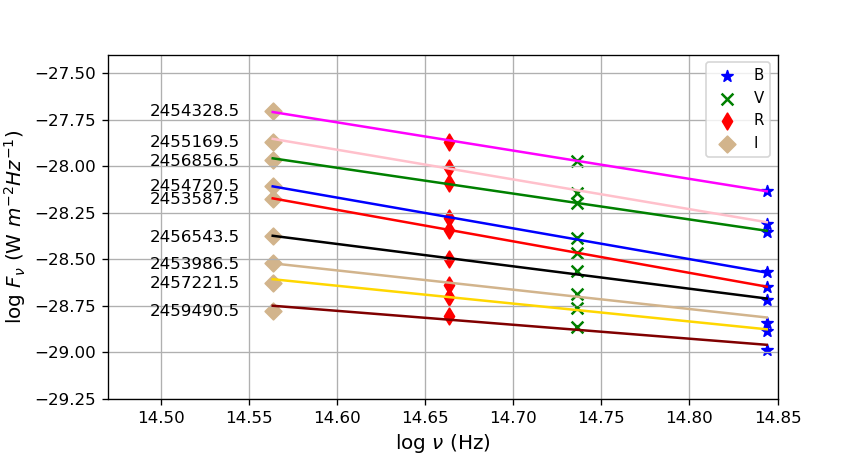}
    \caption{A still image showing nine different spectral indices from the GIF movie. The corresponding Julian dates are shown on the left side, with the legend showing the corresponding band/filter used. The animation runs for a total of 39 seconds showing variation of spectral indices over the observation period of 19 years. The animation provides a visualization of the variation of spectral indices through which the source evolves and shows a nice bimodal distribution that correlates with the high and low states of the source. The animated movie showing the time variation of optical spectral indices for the whole observation period can be found in the online material.}
    \label{fig:5}
\end{figure*}

In this paper, the long-term color variability study was carried out using different pairs of color vs magnitude plots, which revealed a variety of color-magnitude trends through which the object evolved. FSRQ 3C 454.3 has a bright accretion disk, and its multi-wavelength SED is well-fitted using the external Compton model \citep{2012ApJ...758...72W}, which tells us about the presence of soft photons from the broad line emitting region and the dusty torus region. For the object under study, the major emission feature appearing in the spectra is the Mg\RNum{2} line \citep{2019A&A...631A...4N}, which falls at 5200 {\AA} in the observer's frame, between the B and V bands.  Due to the usage of broadband Johnson-Cousin BVRI filters, the intensity of the line is smeared out in the photometric observations and does not contribute significantly to the object's color. Another study conducted by \cite{2017Galax...5....1A} reveals that in bright phases of the source, the contribution of the Mg\RNum{2} line on top of the continuum becomes negligible. 

\begin{figure*}
\centering
    \includegraphics[width=0.60\textwidth]{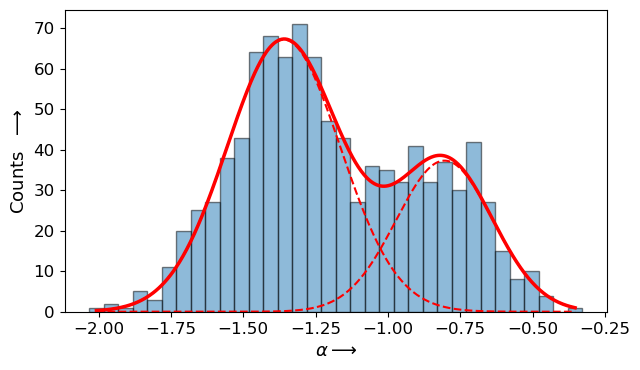}
    \caption{Distribution of spectral indices}
    \label{fig:7}
\end{figure*}

However, since the line does not vary with the same intensity as the continuum, it is worth examining whether the Mg\RNum{2} line will influence the B--V color in the low activity phases. We assume a correlation between the Mg\RNum{2} line and continuum flux for simplicity and that Mg\RNum{2} primarily influences the V-band magnitude, based on the continuum and Mg\RNum{2} line flux retrieved from \autoref{fig:1} in \citet{2019A&A...631A...4N}. We found that the B--V color difference between maximum and minimum continuum flux states, excluding the Mg\RNum{2} contribution, is only 0.25\% larger than that difference when the Mg\RNum{2}  contribution is included. While this estimate may have limitations, it offers a scale for the Mg\RNum{2} line's impact on the B--V color. Given that the B--V colors in this work range from 0.4 to 0.8, this $\sim$0.25\% difference amounts to $\sim$0.01--0.02 and does not significantly alter our following analysis. We can conclude that only two major components dictate the path of the color-magnitude plot, i.e., the jet and the disk emission. Owing to the abundance of nearly simultaneous data points in our study, the interplay of these two components can be nicely seen in \autoref{fig:7} with two well-separated humps, peaking at spectral index values of $\sim -1.36$ and $\sim -0.81$, respectively. The first and the second humps correlate with the bright and faint states of the object, indicating that 3C 454.3 was more commonly in bright phases throughout the observation period. The tail part of the second hump of bimodal distribution approaches zero but never is positive, implying that even in the faintest state of the source, the jet component is present, i.e., the disk emission never entirely swamped the jet emission in any segment. 

The RWB trend indicates that disk emission is more prominent during the object's low flux or quiescent stages \citep{2021ApJ...906....5A}, and in higher flux stages, jet emission dominates. This can be supported by the fact that the object under study is an FSRQ, which are LSPs; hence their first hump peak lies around the IR band, which implies that in the optical regime, the jet emission is \emph{“redder”} in nature, i.e. flux decreases with increasing frequency. On the other hand, the disk emission for black hole masses $\gtrsim$ $10^8$ \(M_\odot\) peaks in the {UV band}; as a result, disk emission will be \emph{“bluer”} in nature, i.e., the flux increases with increasing frequency in the optical regime. Hence, the overall result of the RWB trend indicates that with increasing brightness, the jet emission indeed dominates. Nonetheless, the peak synchrotron frequency need not remain at a particular frequency over such a long observation period. For example, when the jet emission is weak compared to that of the disk, the synchrotron peak frequency can be lower \citep{2013ApJ...763L..11C}. This can cause the spectrum to be much steeper in the optical regime, thereby reducing the redder contribution, which, in effect, will cause the object to be bluer during such a phase. 

One of the studies conducted by \cite{2012ApJ...756...13B} using SMARTS data included 3C 454.3 as a target, and they found that the object followed a RWB trend throughout the observation campaign of $\sim$2.5 years. Another study having 3C 454.3 in the target list, conducted by \cite{2020MNRAS.498.3578S}, found that it displayed RWB behavior throughout the observation campaign of $\sim$ 9 years. Although their plot revealed deviations towards the brighter end, the lack of sufficient data points in the bright state of the object meant they could only fit it using a straight line.

Our long-term study reveals that it is not possible to simply classify this blazar (and probably many others) as RWB or BWB. Rather, the behavior seems to depend on the specific nature of the plasma ejections, the possible interaction of a blob with the quasi-stationary jet component, and other physical processes discussed below, assuming the disk emission varies slowly as compared to the highly fluctuating jet emission. When the jet is in the quiescent phase, or the jet emission has not out-shined the disk emission, the objects trace a localized path on the color-magnitude diagram, remaining in the redder when brighter/bluer when fainter state. But this localization continues only up to a particular magnitude value, shown by a vertical dashed line in \autoref{fig:2}, after which the color plot saturates/flattens and then can redden or become bluer or merely achromatic, which in turn relies upon the processes mentioned above, that are causing the increase in brightness of the source. The flattening that occurs at the break-point results from the canceling of the relative emissions of the jet and the disk, i.e., when their contributions become equal, which is different from the achromatic behavior observed at the bright phases of the object. Up to the point where the RWB trend is strictly followed over the long term, the break-point indicates a crude magnitude value, after which the jet emission completely outshines the disk emission. The RWB continuing after the break-point may be attributed to different electron acceleration events rather than merely an admixture of disk and jet emissions, as the disk component has already been swamped by the jet, signatures of which can be seen in \autoref{fig:2} after the cutoff magnitude. This can happen if the new population of relativistic plasma does not extend as high up in energy as previously cooled ones, causing the spectrum to become softer. The achromatic behavior after jet emission has swamped the disk emission can be explained using diffusive/escape cooling \citep{2017ApJ...844..107I}. If the electron population escapes the emission region much faster (being energy dependent, high energy electrons will leave faster) before they can efficiently radiate, these will not contribute to the color-magnitude plot, causing the object to evolve achromatically in the CM plot. Apart from this, if there is a change in the jet direction and the wavebands get Doppler boosted or de-boosted equally, the blazar will become brighter or fainter, respectively, without any change in the CM plot, provided the spectrum maintains its slope in the Doppler boosted wavebands.

Thus, the evolution of the color-magnitude plot as a source brightens should be treated as a continuous transition from the mixture of jet and disk colors until it flattens and then evolves to BWB, RWB, and achromatic state depending upon the processes involved, causing the change in brightness. It again goes back to disk and jet contributions as the object moves from a bright to a faint state. In effect, the results from previous studies may be regarded as the \emph{“fragments”} of the whole CM plot depending upon the epoch in which the observation occurred. To get the fuller picture, long-term observations with a large number of quasi-simultaneous data points, in both bright and faint states, as we have here for 3C 454.3, are required to properly determine the CM behavior and evolution of other blazars.

\section{Conclusions}
The present work provides the most extensive and densely sampled long-term optical photometric variability studies of the FSRQ blazar 3C 454.3. The data used here was collected from a large number of ground-based optical telescopes around the globe between 2004 and 2023. Below is a summary of our findings: 
\begin{enumerate}
    \item[{1.}] During this lengthy period, this blazar was found to be highly variable, with a change of $\sim$ 5 magnitude in all four (BVRI) optical bands. 3C 454.3 was observed in various flux states: low-states, pre- and post-outburst states, and outburst states on multiple occasions. We divided the total light curve into 19 segments based on the observing season. The minimum and maximum variability amplitudes in a given segment were respectively found to be 0.20 mag in the B band in segment 17 and 4.17 mag in the I band in segment 10.
    \item[{2.}] We studied intraday variability in the R band for this blazar during nine nights when dense observations were available using the nested ANOVA and enhanced F-test. The object was variable on the intraday timescale during six of those nights, as confirmed by the combined results of both tests.
    \item[{3.}] The various color versus magnitude combinations seen in the entire datasets of the BVRI band light curves show clear evidence of a RWB trend, a well-established color versus magnitude trend in FSRQs, but only up to a certain magnitude value. Beyond that magnitude, the CM plot shows saturation and then evolves in a complex fashion, going through BWB, RWB, or achromatic states. These variations may depend on the energetics of the electron population that is causing the change in brightness. Our analysis reveals that it is possible to extract the magnitude value at which jet emission completely outshines the disk emission using an optical CM plot, which is $\sim$ 14.5 in I-band for 3C 454.3, provided sufficient long-term observations are present with a good number of quasi-simultaneous data points in different bands.
    \item[{4.}] We performed cross-correlation analyses in different optical bands for the whole duration of the light curves using the DCF analysis method. The DCF plots in different optical bands all peak at zero lag, indicating that the emission is co-spatial in the BVRI optical bands.
    \item[{5.}] We also generated multi-epoch extinction-corrected optical SEDs for the entire duration of these observations to search for spectral evolution. The optical SEDs are well-fitted by power laws, and we estimated the spectral indices at different epochs during those two decades. The distribution of spectral indices exhibits a bimodal distribution, reflecting the presence of phases when the jet is more dominant or relatively quiescent.
\end{enumerate}

\section*{Acknowledgements}
Based on data taken and assembled by the WEBT collaboration and stored in the WEBT archive at the Osservatorio Astrofisico di Torino – INAF (\url{https://www.oato.inaf.it/blazars/webt/}). This study was based in part on observations conducted using the 1.8m Perkins Telescope Observatory (PTO) in Arizona, which is owned and operated by Boston University. This paper has made use of up-to-date SMARTS optical/near-infrared light curves that are available at {\url{https://www.astro.yale.edu/smarts/glast/home.php}}. Data from the Steward Observatory spectropolarimetric monitoring project were used. This program is supported by Fermi Guest Investigator grants NNX08AW56G, NNX09AU10G, NNX12AO93G, and NNX15AU81G. The ZTF survey is supported by the U.S. National Science Foundation under grants no. AST-1440341 and AST-2034437. The Skinakas Observatory is a collaborative project of the University of Crete, the Foundation for Research and Technology -- Hellas, and the Max-Planck-Institut f$\ddot{u}$r Extraterrestrische Physik. This article is partly based on observations made with the IAC80, the STELLA, and the LCOGT 0.4 m telescopes. The IAC80 telescope is operated by the Instituto de Astrof\'{i}sica de Canarias in the Spanish Observatorio del Teide on the island of Tenerife. Many thanks are due to the IAC support astronomers and telescope operators for supporting the observations at the IAC80 telescope. The STELLA robotic telescopes are an AIP facility jointly operated by AIP and IAC. One of the nodes of the LCOGT 0.4m telescope network is located in the Spanish Observatorio del Teide. \\
\\
CMR, MV, and MIC acknowledge financial support from the INAF Fundamental Research Funding Call 2023. The research at Boston University was supported in part by several NASA Fermi Guest Investigator grants; the latest are 80NSSC22K1571 and 80NSSC23K1507.  RB, ES, and AS were partially supported by the Bulgarian National Science Fund of the Ministry of Education and Science under grant KP-06-H68/4 (2022).  GD, OV, MDJ, and MS acknowledge support from the Astronomical station Vidojevica, funding from the Ministry of Science, Technological Development and Innovation of the Republic of Serbia (contract No. 451-03-66/2024-03/200002), by the EC through project BELISSIMA (call FP7-REGPOT-2010-5, No. 256772), the observing and financial grant support from the Institute of Astronomy and Rozhen NAO BAS through the bilateral SANU-BAN joint research project ``GAIA astrometry and fast variable astronomical objects", and support by the SANU project F-187. Also, this research was supported by the Science Fund of the Republic of Serbia, grant no. 6775, Urban Observatory of Belgrade - UrbObsBel. NRIAG team acknowledges financial support from the Egyptian Science, Technology \& Innovation Funding Authority (STDF) under grant number 45779. The R-band photometric data from the University of Athens Observatory (UOAO) were obtained in the frame of {\it BOSS Project} after utilizing the robotic and remotely controlled instruments at the University of Athens (Gazeas 2016). The IAA-CSIC co-authors acknowledge financial support from the Spanish ``Ministerio de Ciencia e Innovaci\'{o}n" (MCIN/AEI/ 10.13039/501100011033) through the Center of Excellence Severo Ochoa award for the Instituto de Astrof\'{i}sica de Andaluc\'{i}a-CSIC (CEX2021-001131-S), and through grants PID2019-107847RB-C44 and PID2022-139117NB-C44. Some of the data are based on observations collected at the Observatorio de Sierra Nevada, which is owned and operated by the Instituto de Astrof\'{i}sica de Andaluc\'ia (IAA-CSIC), and at the Centro Astron\'{o}mico Hispano en Andalucía (CAHA); which is operated jointly by Junta de Andaluc\'{i}a and Consejo Superior de Investigaciones Cient\'{i}ficas (IAA-CSIC). LC acknowledges the support from the Tianshan Talent Training Program (grant No. 2023TSYCCX0099). 
MFG is supported by the National Science Foundation of China (grant 12473019), the China Manned Space Project with No. CMSCSST-2021-A06, the National SKA Program of China (Grant No. 2022SKA0120102), and the Shanghai Pilot Program for Basic Research-Chinese Academy of Science, Shanghai Branch (JCYJ-SHFY-2021-013). ZZ is funded by the National Science Foundation of China (grant no. 12233005).

%




\bibliography{test}{}
\bibliographystyle{aasjournal}



\end{document}